# Spin-wave dynamics and symmetry breaking in an artificial spin ice

Susmita Saha[1,2,3], Jingyuan Zhou[1,2], Kevin Hofhuis[1,2], Attila Kákay[4], Valerio Scagnoli[1,2], Laura J. Heyderman[1,2], Sebastian Gliga[1,2]

[1]Laboratory for Mesoscopic Systems, Department of Materials, ETH Zurich, 8093 Zurich, Switzerland
[2]Paul Scherrer Institut, 5232 Villigen, Switzerland
[3]Department of Physics and Astronomy, Uppsala University, 751 21 Uppsala, Sweden
[4]Helmholtz-Zentrum Dresden – Rossendorf, Dresden 01328, Germany

Abstract
Artificial spin ices are periodic arrangements of interacting nanomagnets successfully used to investigate emergent phenomena in the presence of geometric frustration. Recently, it has been shown that artificial spin ices can be used as building blocks for creating functional materials, such as magnonic crystals, and support a large number of programmable magnetic states. We investigate the magnetization dynamics in a system exhibiting anisotropic magnetostatic interactions owing to locally broken structural inversion symmetry. We find a rich spin-wave spectrum and investigate its evolution in an external magnetic field. We determine the evolution of individual modes, from building blocks up to larger arrays, highlighting the role of symmetry breaking in defining the mode profiles. Moreover, we demonstrate that the mode spectra exhibit signatures of long-range interactions in the system. These results contribute to the understanding of magnetization dynamics in spin ices beyond the kagome and square ice geometries and are relevant for the realization of reconfigurable magnonic crystals based on spin ices.

Over the past few years, the study of artificial spin ices has evolved from the creation of model systems designed to investigate geometric frustration towards the definition of functional systems.[1-4] Artificial spin ices consist of lithographically patterned, geometrically arranged ensembles of nanomagnets that are magnetostatically-coupled. Each nanomagnet is in a single-domain state with the magnetization pointing in one of two orientations along the magnet long axis due to shape anisotropy[5]. Spin ice geometries, such as the square and kagome ices were originally derived from crystallographic planes in rare earth pyrochlore compounds[6]. Artificial spin ices with these geometries have been found to display rich collective behavior, such as phase transitions[7] and well-defined mode spectra[8-13] based on their large number of micromagnetic degrees of freedom[14-16]. These mode spectra can be tuned by modifying the magnetic state of the system[8] or through coupling with a magnetic underlayer[17] and show promise for creating reconfigurable magnonic crystals[4, 13, 18-20].

Meanwhile, a number of promising geometries have been proposed, beyond the square and kagome lattices. Some enable emergent frustration, rather than geometric frustration at the vertex level[21], leading to control over the dynamics of emergent charges[22, 23], while other geometries have allowed the creation of arrays whose magnetic state is fully reconfigurable[24]. More recently, chiral geometries have been investigated and have been shown to display e.g. ratchet behavior during thermal relaxation[25] and during field-induced magnetization reversal[26] as well as tunable vertex frustration[27], and tunable ferromagnetic and antiferromagnetic order[28-31].

Here, we investigate the magnetization dynamics in arrays as shown in Figure 1a, in which the unit cell is composed of interacting horizontal and vertical nanomagnets, which leads to tunable frustration[27] and anisotropic magnetostatic interactions. While artificial spin ices built upon geometries displaying anisotropic interactions[32] such as chiral structures, have demonstrated rich collective behavior during thermal relaxation[25] and field-induced dynamics[26, 28], their GHz dynamics have not been studied so far. Using time-resolved magneto-optical Kerr effect (MOKE), we determine that the magnetization dynamics is characterized by a rich mode spectrum, and we determine its evolution in an external magnetic field. Finding a complex mode distribution in large arrays, we investigate the evolution of the mode spectrum with system size and geometry, starting with building blocks. We find that the anisotropic magnetostatic interactions contribute to the formation of a large number of modes and that the geometry of the building blocks influences the symmetry of the spatial mode distribution. Surprisingly, while arrays larger than a unit cell exhibit mostly identical mode spectra, irrespective of their geometry, we identify specific modes, whose intensity, and existence, is correlated to the size of the array. The amplitude evolution of these modes reflects the long-range nature of the interactions in the studied system. Our results unveil different mechanisms for mode evolution, which are useful for designing artificial spin ices for applications such as magnonic crystals.

The magnetization dynamics were measured by using an all-optical time-resolved magneto-optical Kerr effect microscope[33, 34] based on two color collinear pump probe technique, shown in Figure 1b. The Kerr rotation of the probe laser beam ($\lambda$ = 1030 nm and pulse width of ca. 50 fs) is measured after exciting the sample by a pump beam ($\lambda$ = 515 nm and pulse width of ca. 50 fs). The pump and probe beams are collinear. The approximate regions covered by the pump and probe beams are schematically indicated in Figure 1a. The pump pulse modifies the local magnetization through thermal effects, thereby inducing precessional magnetization dynamics. A Fourier transform of the time resolved data is performed to obtain the mode spectra of the sample at different values of an external magnetic field. This field is applied at 45° in the sample plane (as indicated in Fig. 1a) in order to maintain an equal contribution of different sublattices of the system (consisting of the vertical and horizontal nanomagnet groups). The measured spin-wave spectra in the presence of different applied fields are shown in Fig. 2a (See Supplementary Information Fig. S1 for the experimental time traces and Methods for the details of the measurements). For comparison, in Fig. 2b we plot the results of micromagnetic simulations, where the simulated FFT is performed considering a region of the array equivalent to the probe laser spot (See Methods for details about the micromagnetic simulations). Distinct modifications of the spectra are visible, which are reproduced by the simulations: in particular, the red shift of the peak with the largest amplitude, along with the splitting of this peak as the applied field is reduced. The relative discrepancies in intensities and the precise positions of the peaks in the frequency domain are not always reproduced owing to the presence of a variety of effects, e.g. edge roughness, and the possible difference in the saturation magnetization and gyromagnetic ratio values in the experimental samples. The mode evolution as a function of magnetic field strength is simulated in Fig. 2c, demonstrating how the mode spectra (frequency and amplitude) are modified. The field-dependent spectra are the result of changes in the magnetic structure: at 100 mT, the nanomagnets are almost uniformly magnetized along the applied field direction, while at 30 mT, the magnetization is mostly aligned along the long axis of the nanomagnets, displaying a curling of the magnetization at the element edges[14, 15, 26]. Consequently, by applying a sufficiently large field (e.g. 100 mT) along the diagonal direction, the magnetization has a significant component normal to the edges of the nanomagnets, such that the interactions between the horizontal and vertical nanomagnets is mediated by surface charges, $\sigma \propto \boldsymbol{m} \cdot \boldsymbol{n}$, where $\boldsymbol{m}$ is the magnetization vector and $\boldsymbol{n}$ a vector normal to the surface. At lower fields (e.g. 30 mT), the interactions are mediated by surface charges as well as by volume charges, $\rho \propto \boldsymbol{\nabla} \cdot \boldsymbol{m}$, located at the extremities of the nanomagnets. Further insight into the nature of the modes can be obtained by considering their spatial distribution. In Figure S2 of the Supplementary Information, we plot the simulated mode distributions for a

5x5 array at 100 mT, which exhibit complex spatial profiles, especially above ca. 8 GHz. Owing to their complexity, these modes cannot be easily analyzed. Nevertheless, we can obtain an understanding of their origin by considering building blocks of the array.

The spin-wave dynamics of three parallel nanomagnets, a 2x1 asymmetric structure and the 2x2 structure are simulated to determine the dependence of the mode spectra on geometry (Figure 3a). While the spectra display similar features, they become richer due to the increasing number of modes as the number of nanomagnets increases and their geometry is modified. However, modes are often very close in frequency and the finite damping results in overlap of the peaks. The modes below 8 GHz correspond to edge modes where the magnetization oscillates at the edges of the nanomagnets (see Supplementary Information Fig. 2 at 6.5 GHz), while the modes above 12 GHz correspond to modes quantized along the length and width of the nanomagnets (see Supplementary Information Fig. 2 for the mode profile at 15 GHz). Because both these types of modes are rather weak and, in practice, difficult to measure (except possibly with micro-focused Brillouin Light Scattering[35]), we concentrate here on the modes in the range 8 GHz – 12 GHz, and in particular on the few modes plotted in Fig. 3b-d, whose evolution we trace as a function of the geometric configuration of the nanoislands. In the three-magnet horizontal and vertical structures (Fig. 3b), the mode at 9 GHz (highlighted by a red dot) corresponds to modes mostly localized in the outermost elements and quantized along the nanomagnet length. Interestingly, similar modes have the same frequency in the 2x1 structure configuration (where the horizontal and vertical elements form a magnetostatically coupled system) as well as in the 2x2 structure that corresponds to the same edge oscillations of outermost elements. The spatial distribution of these modes reflects, in part, the symmetry of the systems. The three-magnet configurations display inversion symmetry[36] about the central element (labeled $X_2$ and $Y_2$), and so does the mode distribution. The asymmetry of the 2x1 structure leads to a different mode distribution, in which the edges only oscillate in elements labeled $X^T_1$ and $Y^T_3$. While the global inversion symmetry is restored in the 2x2 structure and the mode distribution is overall symmetric, but only elements $Y^{2x2}_1$ and $Y^{2x2}_6$ display pronounced edge oscillations. The lack of oscillations in the geometrically equivalent elements $X^{2x2}_3$ and $X^{2x2}_4$ is a result of the applied field, $\mu_0 H = 100$ mT, which leads to further symmetry breaking. In contrast, the mode at 9.92 GHz (second row of Fig. 3 b-d, highlighted by blue rectangles) is slightly blue-shifted and its relative amplitude decreases in as the geometry of the structure becomes more complex. In the three-magnet state, it corresponds to quantized modes in the outermost nanomagnets that display inversion symmetry, while the mode distribution in the central nanomagnet is mostly bulk-like. In the 2x1 structure, a mode with similar features occurs at 9.96 GHz (purple square), where the mode distribution in element $Y^T_3$ is similar to that in $Y_3$ and the mode distribution in $X^T_1$ is similar to that in $X_1$. Element $Y^T_2$ exhibits a mode distribution very similar to that found in $Y_2$ and in $X_2$. The other elements display fundamentally different oscillations. Surprisingly, similar modes are found at 10.12 GHz in central elements of the 2x2 structure: the magnetization oscillations in elements $X^{2x2}_1$ and $X^{2x2}_6$ are the same as in $X_1$ and $X_3$, respectively. The (outer) elements $X^{2x2}_3$ and $X^{2x2}_4$ equally display similar modes. Moreover, the mode distribution in element $X^T_3$ (2x1 structure) at 9.96 GHz is also found at 9.96 GHz in elements $X^{2x2}_3$ and $X^{2x2}_4$ of the 2x2 structure. Features of the mode at 9.96 GHz in the three-magnet geometries are thus split over multiple modes in the 2x2 structure. Finally, we consider the mode at 10.08 GHz in the 2x1 structure (highlighted by the green triangle), which has no equivalent in the three-magnet state. Features of that mode are present in the 2x2 structure. In particular, the mode distribution in $X^{2x2}_1$ and $X^{2x2}_6$ reflects that of $X^T_1$ and $Y^T_3$, respectively, while the oscillations in $Y^{2x2}_4$ have the same features as in $Y^T_1$. The oscillations of the magnetization in $X^{2x2}_3$ and $X^{2x2}_4$ also are similar to those in $X^T_1$ and $Y^T_3$ (in terms of the number of nodes and symmetry).

The observed spatial mode distributions are governed by the demagnetizing field associated with the element ensembles, which are plotted in Fig. 4. The demagnetizing field in element $Y_3$ of the three-magnet structure (Fig. 4a) and element $Y^T_3$ in the 2x1 structure (Fig. 4b) have very similar demagnetizing fields, leading to the similar mode profiles described in Fig. 3b. The demagnetizing field in the other elements of the 2x1 structure is, however, modified by the magnetostatic interactions between the elements, in particular in elements $X^T_{2,3}$ and $Y^T_{1,2}$, with the most substantial modifications occurring in $Y^T_1$, which strongly interacts with all three horizontal nanomagnets. These interactions break the inversion symmetry observed in the demagnetizing fields of the elements within the three-magnet structure (Fig. 4a). In addition, while the demagnetization field within $Y^T_2$ is mostly similar to that in $Y_2$, a slight asymmetry is present, which explains why the mode distribution in $Y^T_2$ at 9.0 GHz and at 9.96 GHz slightly differs from that in $Y_2$. In the 2x2 structure, the large number of magnetostatic interactions leads to a different distribution of the demagnetizing field (Fig. 4c). In particular, the demagnetizing field profiles in $X^T_3$ and $X^{2\times2}_3$ are similar, resulting in the similar mode profiles at 9.96 GHz (Fig. 3b-c).

Having investigated the modes of configurations up to the 2x2 structure, we now consider different sizes of nanomagnet arrays. In Supplementary Information Fig. S3, mode spectra are plotted for configurations ranging from a 2x2 structure to a 5x5 array of three-magnet groups. The eigenmodes are similar, independent of the array size and edge configuration, indicating that the dominant modes are contained within the 2x2 structure, which thus constitutes the magnetic unit cell, establishing that the studied pattern has a chiral unit cell. Interestingly, a number of modes display an amplitude dependence on array size. In particular, two peaks plotted in Fig. 5a (at 10.88 GHz and 11.04 GHz) have a distinct evolution as a function of the array size. Fig. 5b shows that the amplitude of the mode at 10.88 GHz decreases with increasing array size. Plotting the spatial mode profile in Fig. 5c demonstrates that the mode amplitude is mostly localized within nanomagnets around the perimeter of the array (highlighted by the blue region). Fitting the data in Fig. 5b indicates that the mode amplitude decay follows an inverse square law, indicating the presence of significant long-range interactions in the system, in agreement with the observation in Ref. [27], where long-range ordered ground-state configurations have been observed. This is in contrast to the square ice, where mode amplitudes display a linear evolution[8], due to the predominance of nearest-neighbor interactions[37]. In parallel, the amplitude of the mode at 11.04 GHz increases with increasing array size because the mode profile displays significant amplitude within the bulk of the array (Fig. 5d and Supplementary Information Fig. S4). Fitting the data indicates a square root-like behavior, starting to level off beyond the 5x5 array.

Conclusion

We have determined the magnetization dynamics in a spin ice geometry displaying anisotropic magnetostatic interactions, which give rise to a rich mode spectrum. Using time-resolved magneto-optical Kerr microscopy, we find that an external magnetic field can significantly modify the spin-wave spectrum. In addition, micromagnetic simulations show that the spatial distribution of the resulting modes strongly depends on the local symmetry and distribution of the demagnetizing field. The structural asymmetry of the system, in which groups of vertical nanomagnets interact with horizontal ones, leads to a proliferation of modes with complex spatial distributions. While most modes are present in the 2x2 unit cell of the array, we have found that the amplitude of certain well-defined modes varies as a function of the array size: depending on whether they are 'edge' or 'bulk' modes of the array, their amplitude will respectively decrease or increase, and thus indicate the presence of long-range interactions within the system. Interestingly, a distinct mode appears around 11 GHz, whose amplitude increases with array size, providing a means of determining the array size based on the mode spectrum. These results indicate how symmetry breaking can be exploited to tune the resonant spectrum of artificial spin ices. This spectral tunability opens the way to the use of artificial spin ices as reconfigurable magnonic crystals[4, 19, 20, 38]

The Supporting Information is available free of charge on the ACS Publications website at DOI: XXXX.

Figures:

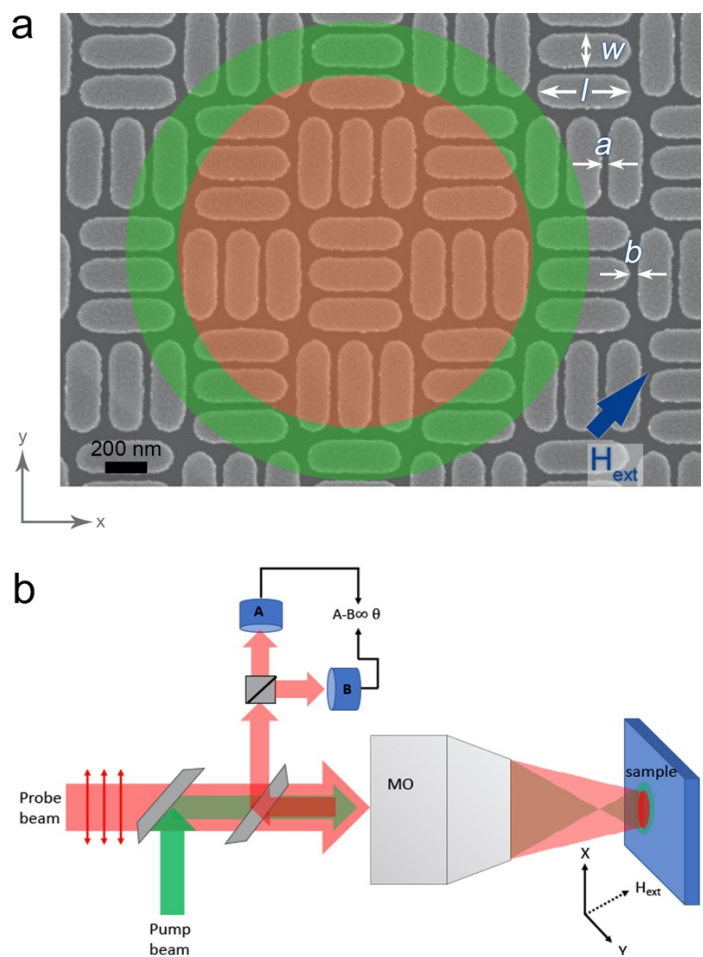

Fig. 1 (a) Scanning electron microscopy (SEM) image of the Permalloy ($Ni_{83}Fe_{17}$) spin ice lattice, with the geometry of the applied magnetic field. The edge-to-edge separation between individual nanomagnets is labeled *a* and the separation between groups of three nanomagnets is labeled *b*. In this case, *a*= 50 nm and *b*= 50 nm. The approximate excitation region of the pump laser beam is indicated by a green ellipse, whereas the red ellipse represents the average probed region. (b) Schematic of the time resolved magneto-optical Kerr microscope setup. A linearly polarized probe beam is used to measure the magnetization dynamics after exciting the sample using a pump beam. Both pump and probe beam are focused on the sample using a microscope objective (MO). A magnetic field ($H_{ext}$) is applied at 45° to the sample plane. The precessional dynamics is measured from the reflected beam using a balanced photodetector with two photodiodes represented by A and B.

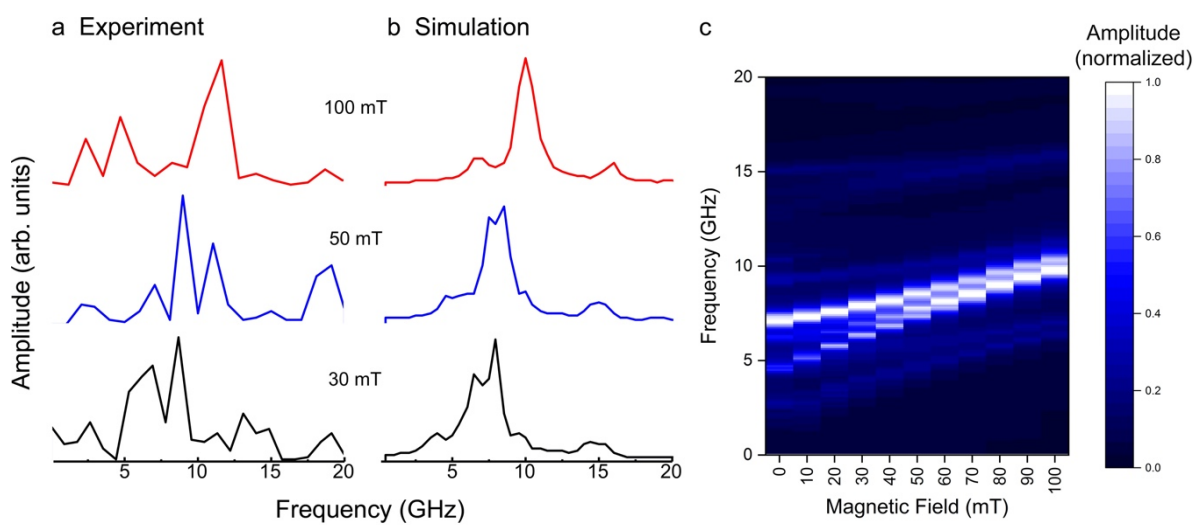

Fig. 2. Spectra of the (a) experimental and (b) simulated magnetization modes considering two nanoseconds relaxation time and a limited excited region, as sketched in Fig. 1a, for three different magnetic field strengths: $\mu_0 H_{ext}$=30 mT, 50 mT and 100 mT. While individual modes cannot be cross-correlated, the general distribution of the simulated modes agrees with the experimentally measured ones. (c) Simulated external magnetic field dependence of the spin-wave spectra.

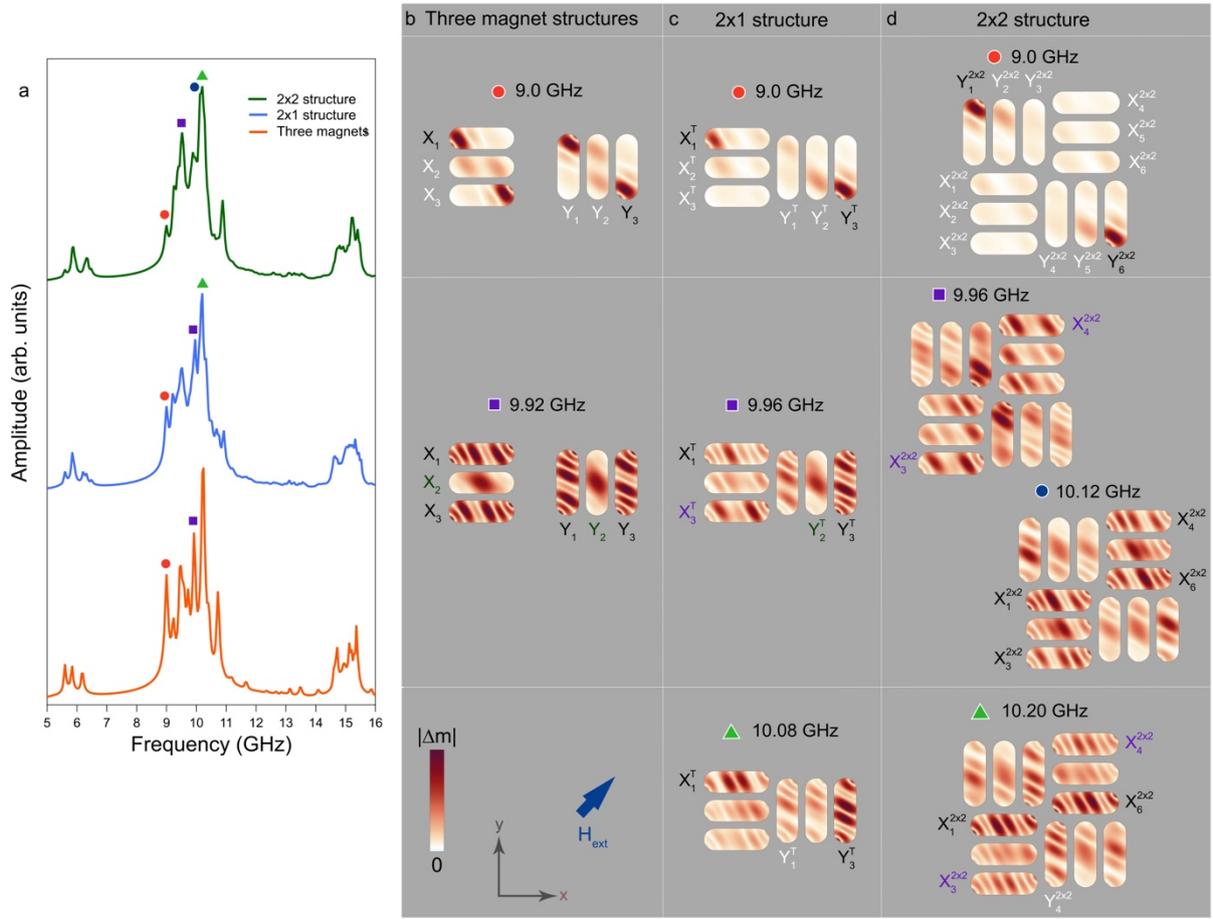

Fig. 3: (a) Simulated spinwave spectra of three parallel nanomagnets, a 2x1 configuration and of a 2x2 magnetic unit cell with $\mu_0 H_{ext}$=100 mT applied along the direction indicated by the blue arrow. The spatial profile of the labeled modes are plotted in (b), (c) and (d) for different building block geometries. The modes are identified based on their spatial distribution. The labels ($X_1$, $Y_1$, etc.) are indicated for all nanomagnets in the top row. Subsequently, only nanomagnets of interest are labeled. Black, blue and green labels highlight similar modes across a row. White labels are general island labels in the top row. The modes are marked on the spectra and in (b) – (d) by corresponding colored circles, rectangles and triangles.

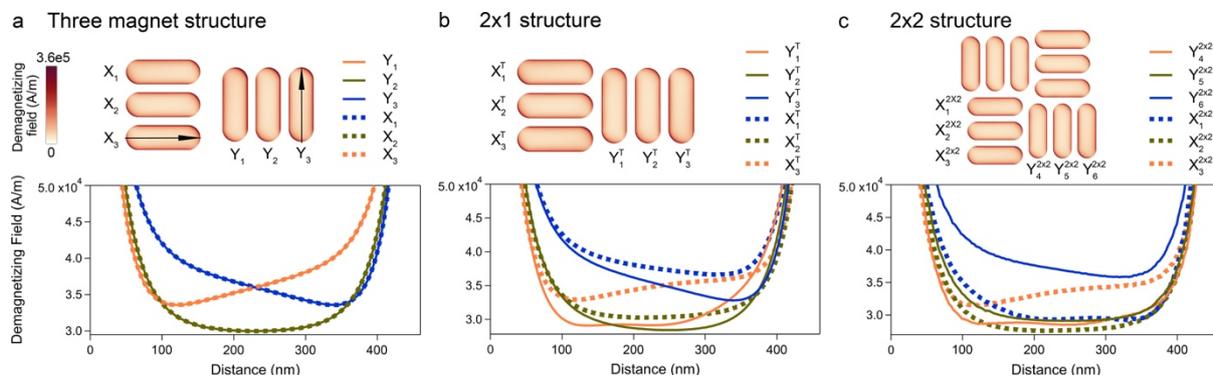

Fig. 4 Simulated demagnetization fields for (a) three horizontal and three vertical magnet structures, (b) a 2x1 structure and (c) and 2x2 structure with $\mu_0 H_{ext}$=100 mT. The demagnetizing field is plotted along the arrow directions indicated in (a). The three magnet structure clearly displays inversion symmetry, which is broken in the T structure.

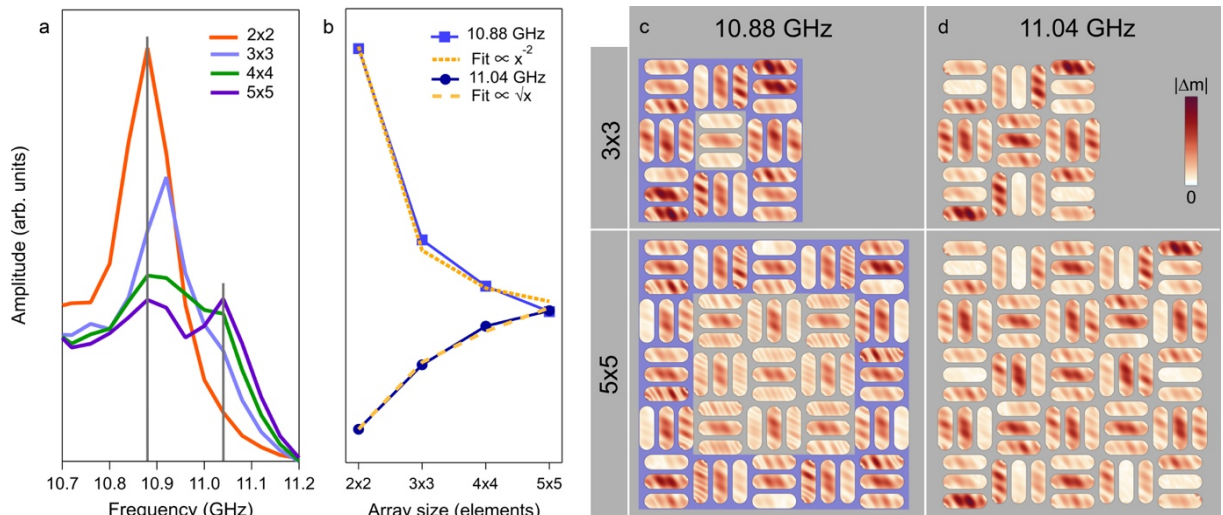

Fig. 5 (a) Mode spectra, highlighting the modes at 10.88 GHz and 11.04 GHz (indicated by the grey lines), for different square array sizes ranging between 2x2 and 5x5 groups of vertical and horizontal three-element structures. (b) Evolution of the mode amplitudes as a function of array size, with fits. (c), (d) Spatial distribution of the mode amplitudes: the power is mostly concentrated in groups of elements along the perimeter of the array (highlighted by the blue region) for the mode at 10.88 GHz, while it is distributed across elements within the array at 11.04 GHz.

## Methods

**Sample preparation**

Finite arrays of Permalloy ($Ni_{83}Fe_{17}$) nanomagnets were prepared on a silicon (100) substrate using electron beam lithography in conjunction with thermal evaporation at room temperature and a base pressure of $2 \times 10^{-7}$ mbar followed by lift-off. The evaporation resulted in a nanocrystalline Permalloy film, which was capped by a 3 nm aluminium layer to protect against oxidation. The nanomagnets are 450 nm long, 150 nm wide and 20 nm thick. The edge-to-edge separation between the individual nanomagnets is 50 nm whereas the edge-to-edge separation between two trident building blocks are 50 nm.

**MOKE measurements**

The magnetization dynamics was measured using a two colour optical pump-probe technique. The second harmonic ($\lambda$ = 515 nm, pulse width ≈50 fs) of a Fiber pulsed laser of wavelength $\lambda$ = 1030 nm and pulse width ≈50 fs is used to excite the sample. The fundamental laser beam ($\lambda$ = 1030 nm) is used to probe the dynamics after passing through a variable time delay by measuring the polar Kerr rotation using a balanced photo diode detector. Both the pump and probe beams are made collinear and are focused by using a microscope objective of Numerical aperture N.A. = 0.65. The polar Kerr rotation of the back reflected beam is measured by means of a balanced photodiode detector. The pump beam is slightly more defocused than the probe beam, as shown in Fig 1, which makes it easier to overlap both the pump and probe beam on the sample surface. The pump beam diameter is of ca. 3 μm. The applied magnetic field is tilted slightly out of the plane of the sample to have a finite demagnetizing field along the direction of the pump pulse. The precessional dynamics consists of an oscillatory signal on top of the exponentially decaying time resolved Kerr rotation. A fast Fourier transform (FFT) is performed after subtracting the bi-exponential background to determine the corresponding power spectra. The measurement time window used in this experiment is 2 ns.

**Micromagnetic simulations**

Micromagnetic simulations based on the Landau-Lifshitz-Gilbert (LLG) equation were performed using *mumax$^3$* [39] and TetraMag[40]. The parameters used for the simulations were: damping $\alpha$ = 0.006[41], exchange constant, $A$ = 1.3 x 10$^{-11}$ J/m, magnetocrystalline anisotropy constant $K$ = 0, and saturation polarization $\mu_0 M_s$ = 1 T. The sample geometries and sizes used in the simulations are the same as in the experiment, as determined from SEM images. The static external field, $\mu_0 M_{ext}$, is applied according to the experimental configuration. The excitation field is a *sinc* function applied perpendicular to the sample plane. For the simulated data shown in Fig. 2b, the field is applied over a region comparable to that excited by the laser in the experiments (see Fig. 1a) and the magnetization dynamics integrated for 2 ns using MuMax. The the simulated data shown in Fig. 2c, the dynamics is integrated for 10 ns. In all other figures, the magnetization was integrated for 50 ns using TetraMag. The mode spectra and the spatial profiles of the modes are obtained by calculating the Fourier transform for each cell. In MuMax, the samples are discretized into prism-like cells with dimensions of 2 x 2 x 20 nm$^3$. In TetraMag, the samples are discretized in tetrahedral elements with a side length of ca. 5 nm.

Acknowledgements:

S.S. acknowledges ETH Zurich Post-Doctoral fellowship and Marie Curie actions for People COFUND program (Grant No. FEL-11 16-1). J.Z. and K.H. acknowledge financial support from the Swiss National Science Foundation (Project Number: 200020_172774). S.G. acknowledges funding from the Swiss National Science Foundation, Spark project no. 190736. S.S. and S.G.

are thankful to Alan Farhan for fruitful discussions. S.G. thanks Simone Finizio for assistance with the three-dimensional data plotting. Use of the Center for Nanoscale Materials, an Office of Science user facility, was supported by the US Department of Energy, Office of Science, Office of Basic Energy Sciences, under Contract No. DE-AC02-06CH11357.